\documentclass[aps,prb,twocolumn,graphicx]{revtex4-1}

\usepackage{graphicx}

\begin{document}

\title{Stability of exfoliated Bi$_2$Sr$_2$Dy$_x$Ca$_{1-x}$Cu$_2$O$_{8+\delta}$ studied by Raman microscopy} 
\author{L.J. Sandilands}
\author{J.X. Shen}
\author{G.M. Chugunov}
\author{S.Y.F. Zhao}
\affiliation{Department of Physics \& Institute for Optical Sciences, University of Toronto, 60 St. George St., Toronto, Ontario M5S 1A7, Canada}
\author{Shimpei Ono}
\affiliation{Central Research Institute of Electric Power Industry, Yokosuka, Kanawaga 240-0196, Japan}
\author{Yoichi Ando}
 \affiliation{Institute of Scientific and Industrial Research, Osaka University, Ibaraki, Osaka 567-0047, Japan}
\author{K.S. Burch}
\affiliation{Department of Physics \& Institute of Optical Sciences, University of Toronto, 60 St. George Street, Toronto, ON M5S 1A7}

\begin{abstract}
Nearly nanometer thick cuprates are an appealing platform for devices as well as exploring the roles of dimensionality, disorder, and free carrier density in these compounds. To this end we have produced exfoliated crystals of Bi$_2$Sr$_2$Dy$_x$Ca$_{1-x}$Cu$_2$O$_{8+\delta}$ on oxidized silicon substrates. The exfoliated crystals were characterized via Atomic Force and polarized Raman microscopies. Proper procedures for production, handling and monitoring of these thin oxides are described. We observe a significant change in the effective exchange constant $J$ of these exfoliated crystals.
\end{abstract}


\maketitle 
\section{Introduction}
Cuprate thin films are a promising avenue for exploring issues of dimensionality, disorder, and free carrier density in these strongly correlated materials\cite{ahn,Konstantinovic2000859,ahn1999, PhysRevB.51.3257}. Mechanical exfoliation has proven to be an effective way of producing a variety of thin crystals\cite{geim2005,staley:184505,top,Ye:2010fk}, although it has mostly been applied to graphene. Compared to traditional techniques, mechanical exfoliation allows a large variety of deposited materials and substrates. To determine the viability of this approach in the high-temperature superconductors,  we have produced thin cuprate crystals on oxidized Si substrates via mechanical exfoliation and characterized them with atomic force microscopy (AFM) and micro-Raman spectroscopy. The measured Raman spectra of these exfoliated samples suggest differences in the magnetic properties versus bulk and provide a useful tool for monitoring sample quality. 

\section{Experiment}
\subsection{Exfoliation Methods and Microscopy}
Heavily under-doped  Bi$_2$Sr$_2$Ca$_{1-x}$Dy$_x$Cu$_2$O$_{8+\delta}$ (Bi-2212) crystals (x = 0.3,0.4) were grown by the floating zone method and characterized by SQUID magnetometry, resistivity, Hall, and thermal transport measurements.\cite{PhysRevB.77.094515} These samples are highly insulating and exhibit no magnetic or superconducting order down to 2 K. Thin Bi-2212 samples were deposited onto Si substrates capped with a 280 nm thick SiO$_2$ layer via an exfoliation technique similar to those applied to graphene\cite{geim2005}. To minimize sample exposure to adsorbed water and air, the substrates were baked in N$_2$ gas at 400 K prior to deposition. Next the substrates are allowed to cool to room temperature, and then exfoliation is performed in the same N$_2$ environment. Upon deposition, exfoliated crystals are located using an optical microscope. The reflectivity of these samples is sensitive to thickness, allowing the  identification of thin crystals. Indeed, we have identified crystals as thin as two unit cells ($\approx 6$ nm) in this manner. Finally, the samples are stored in vacuum when not in use.

\begin{figure}
\includegraphics[scale=1]{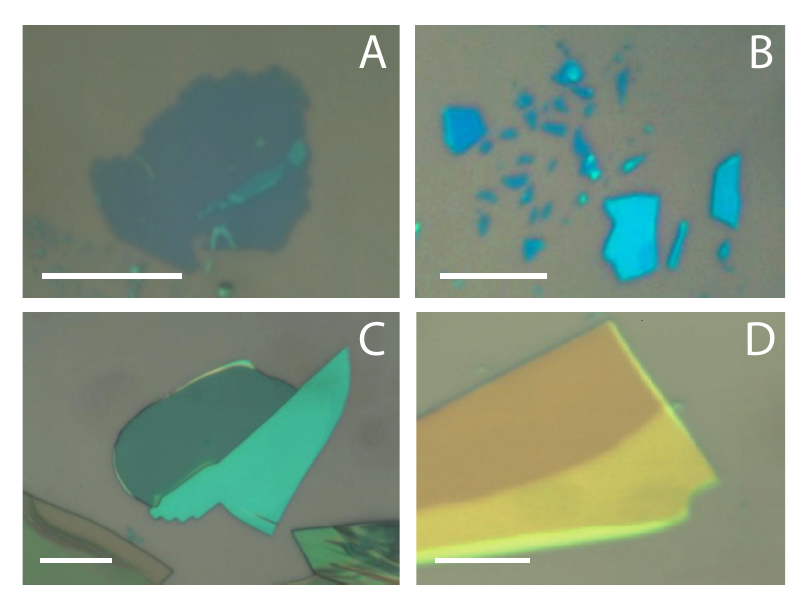}
 \caption{\label{fig:flakes} Optical microscope images of exfoliated crystals. All scale bars are 20 microns. A) 13 nm thick crystal (dark cyan) B) 17 nm (dark blue) and 20 nm (light blue) thick crystals C) 42 nm thick crystal (bright green region) and 160 nm thick crystal (dark green region) D) Exfoliated crystal of varying height. The yellow and orange regions are both thicker than 100 nm.}%
 \end{figure}
\begin{figure}
 \includegraphics[scale=1]{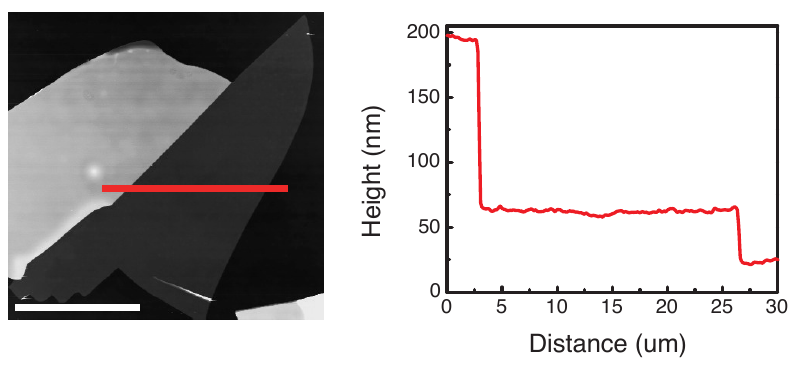}%
 \caption{\label{fig:afm2} AFM image (left panel) and profile (right panel) of a 42 nm exfoliated crystal. The red (gray) line in the image corresponds to the profile displayed in the right panel. An optical image of this same crystal is shown in figure \ref{fig:flakes} C).} %
 \end{figure}

Example exfoliated crystals of a variety of thicknesses (and thus colors) are shown in figure \ref{fig:flakes}. After being identified optically, potentially thin crystals were then examined using a Digital Instruments Nanoscope III AFM operating in contact mode to measure their dimensions. An example AFM image is shown in figure \ref{fig:afm2}. These exfoliated crystals can be tens of microns across and as thin as three unit cells. As seen in figure \ref{fig:afm2}, samples tend to be planar (RMS roughness $<$ 1.5 nm comparable to the bare substrate roughness) and have well-defined edges. The small thickness and large cross-sectional area of these samples makes them promising candidates for studies of electrostatic doping and the effects of dimensionality. 

\begin{figure}
 \includegraphics[scale=1]{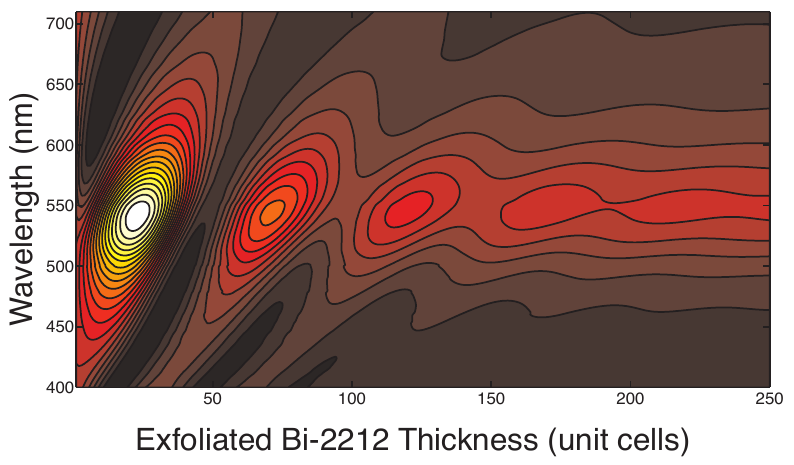}%
 \caption{\label{fig:visibility}Calculated exfoliated crystal visibility versus wavelength and crystal thickness on a 280 nm SiO$_2$/Si substrate.} %
 \end{figure}

Our AFM and optical microscopy studies allow us to explain the observed color variation with height. This behavior can be understood as an interference effect. Specfically, contrast can be defined as the difference in reflected intensity from an exfoliated crystal and the bare substrate normalized to the substrate value. Using bulk optical constants from reference \onlinecite{PhysRevB.60.14917}, a calculation of crystal contrast similar to reference \onlinecite{blake:063124} suggests that the peak contrast should blue-shift with decreasing thickness. The results of this calculation are shown in figure \ref{fig:visibility} where we display the contrast as a function of crystal thickness and wavelength. The contrast for a 20 unit cell crystal peaks in the green. As the crystals become thinner, the peak contrast blue-shifts and decreases.
 \begin{figure}
 \includegraphics[scale=1]{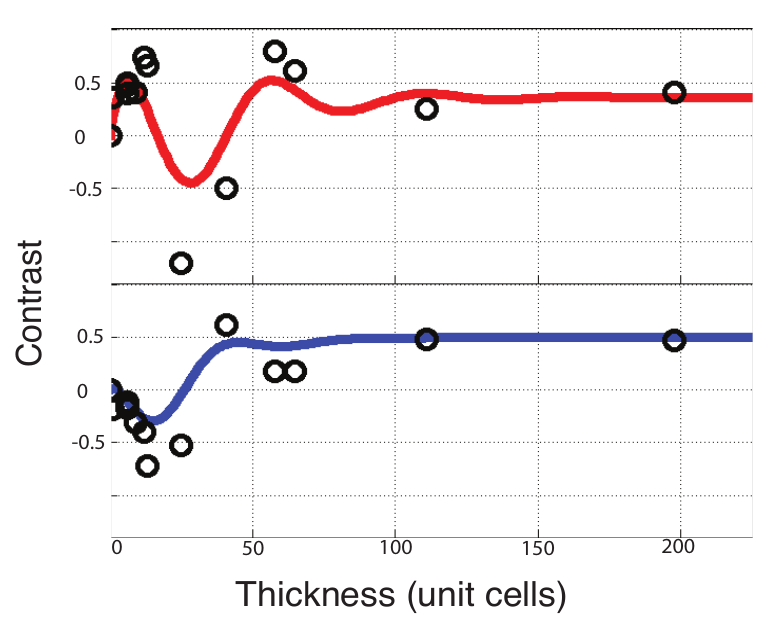}%
 \caption{\label{fig:grig} CCD visibility measurements for red (top) and blue (bottom) channels. The solid curves represent the visibility expected from the calculation illustrated in figure \ref{fig:visibility} while open circles are the measured crystals visibilities.} %
\end{figure}
To confirm this calculation, visbility measurements were performed using a color CCD camera coupled to an optical microscope. Figure \ref{fig:grig} displays the measured (open circles) and calculated (curves) crystal visibilities for a variety of heights. The calculated visibilities were determined by integrating the calculated contrast over the spectral response of a given CCD channel. These measurements show the results of the contrast calculation to be qualitatively correct.  While this data suggests that these exfoliated pieces are indeed Bi-2212, it is necessary to establish whether these exfoliated crystals remain Bi-2212 or form some other phase. 

\subsection{Raman Spectroscopy}
Polarized Raman spectroscopy is a well established tool for identifying materials and has been widely applied to the cuprates\cite{weber_merlin,PhysRevB.68.184504,devereaux:175,PhysRevB.46.6505,PhysRevB.37.2353,PhysRevB.42.8760,PhysRevB.43.3009,PhysRevB.42.4842,PhysRevB.61.9752,Weber:89}. A Horiba Jobin Yvon LabRam Raman microscope with a 532 nm excitation source and 100x (0.8 NA) microscope objective was used to measure the Raman spectra of bulk and exfoliated samples. Measurements were performed in the backscattering geometry. The polarizations of the incident and scattered light are defined with respect to the sample crystallographic axes. The notation XY, for instance, refers to incident light polarized along the $a$ axis (along the Cu-O bond) with scattered light polarized along the $b$ axis; while X' and Y' refer to directions 45 degrees from X and Y respectively. Raman spectra contain components corresponding to the different symmetry projections of the material and these components can be distinguished by their different polarization dependences. Dy-doped Bi-2212 is known to be nearly tetragonal and so we take spectra in the XX, X'X', XY, and X'Y' polarization geometries. For tetragonal (D$_{4h}$) crystal symmetry, XX corresponds to B$_{1g}$ + A$_{1g}$, X'X' to B$_{2g}$ + A$_{1g}$, XY to B$_{2g}$ + A$_{2g}$, and X'Y' to B$_{1g}$ + A$_{2g}$.\cite{devereaux:175} At Raman shifts less than 1000 cm$^{-1}$  exfoliated sample spectra are dominated by signal from the silicon substrate, making phonon peaks difficult to identify. Fortunately, bulk Bi-2212 has two features at higher energies: a collection of two-phonon peaks broadened into a single feature around 1250 cm$^{-1}$ and a broad two-magnon feature centered between 2000-3000 cm$^{-1}$ (depending on doping) \cite{weber_merlin,PhysRevB.68.184504,G.Blumberg11211997,PhysRevB.53.8619}. The two-magnon feature has been studied extensively due to the importance of magnetism in cuprates physics\cite{fluc}. Since thin materials are easily damaged by laser radiation\cite{staley:184505}, we tried a variety of exposure times and excitation powers to maximize the signal to noise ratio while minimizing exposure to the Raman excitation source. Limiting the samples exposure to the laser to approximately ten seconds at 0.5 mW produced no observable changes in the Raman response. Finally, the measured spectra have been corrected for interference effects as described elsewhere\cite{yoon:125422}.

 \begin{figure}
 \includegraphics[scale=1]{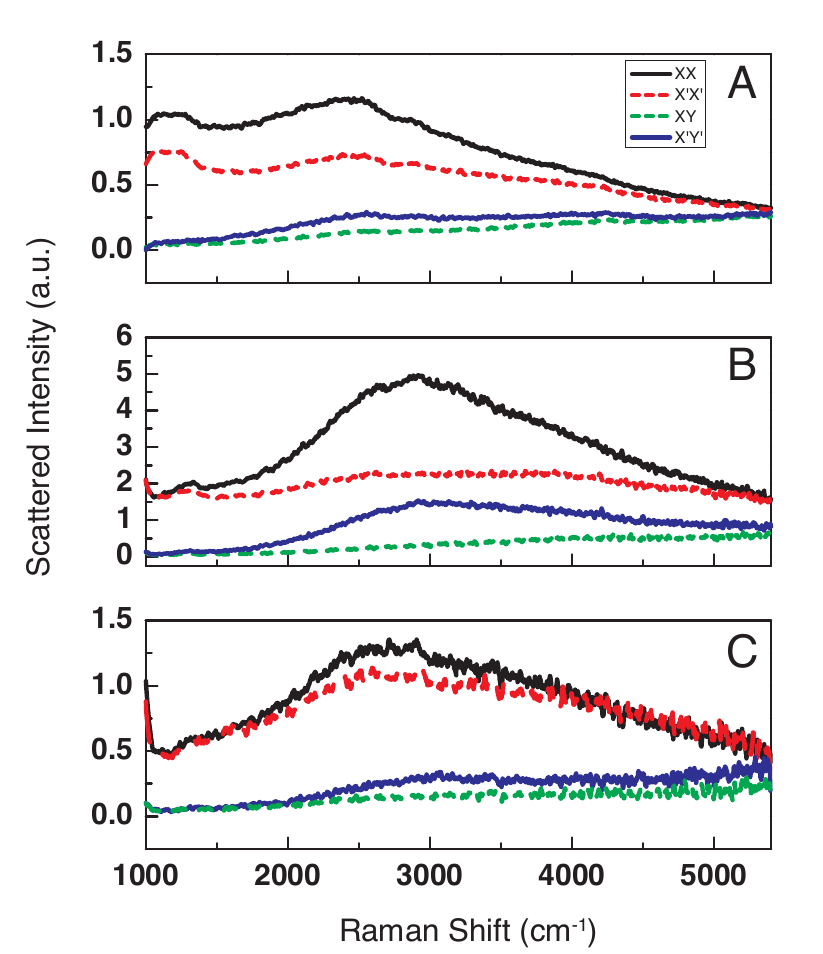}%
 \caption{\label{fig:raman} Polarized Raman spectra for a) bulk b) 42 nm exfoliated crystal c) degraded 30 nm exfoliated crystal. In each case, XX spectra are shown in solid black, X'X' in dashed red (dark gray), XY in dashed green (light gray), and X'Y' in solid blue (dark gray).    XX and X'X' spectra have been offset for clarity. The spectra in a) and b) show a marked polarization dependence compared with c).} %
 \end{figure}
 
The bulk spectra shown in figure \ref{fig:raman}a show both the two-phonon and the two-magnon features seen in previous experiments. Raman intensity units are arbitrary but consistent between plots. In agreement with previous studies, the two-phonon feature near 1300 cm$^{-1}$ appears primarily in the fully symmetric A$_{1g}$ channel (XX and X'X'). In addition, a broad two-magnon feature appears in B$_{1g}$ (XX and X'Y') near 2500 cm$^{-1}$. For the most part, these two features are also visible in the exfoliated samples. Note that only data for x = 0.4 crystals is displayed, although the behavior of x = 0.3 bulk and exfoliated crystals are qualitatively identical. The Raman features observed in typical exfoliated crystals (figure \ref{fig:raman}b) generally retain the selection rules expected from bulk measurements and from the literature for an $ab$-face. This confirms that the cleavage plane is in fact the $ab$-plane, as observed in other experiments\cite{Tanaka:1989uq}. In both exfoliated and bulk samples,we find that the two-magnon feature is principally in the B$_{1g}$ symmetry channel while the two phonon excitation appears predominantly in A$_{1g}$. This suggests that exfoliated samples retain the crystal symmetry of the bulk. The polarization dependence of the two-magnon feature in Bi-2212 also allows the identification of the crystal axes of an exfoliated sample, although it cannot distinguish between the $a$ and $b$ axes. 

\begin{figure}
 \includegraphics[scale=1]{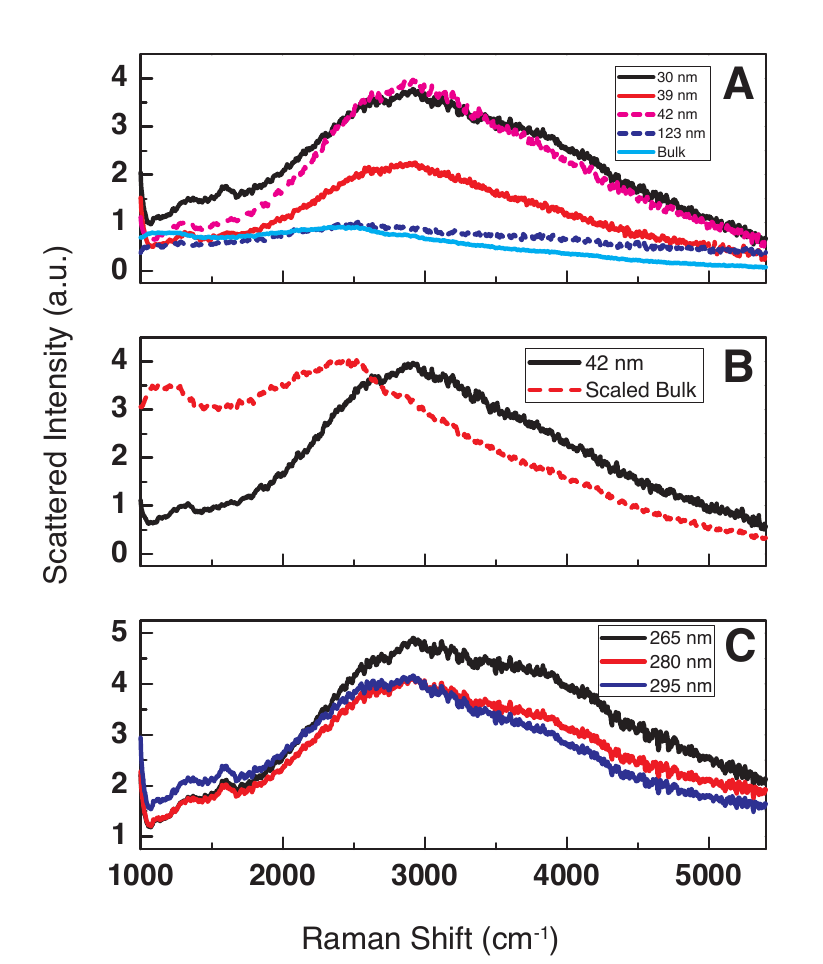}%
 \caption{\label{fig:disc} a) Variation of XX spectra with thicknesses. b) XX spectra for bulk and a 10 unit cell exfoliated sample showing a shift in the two-magnon peak position. Data has been normalized by the two-magnon peak height. c) Effect of oxide layer variation on interference corrected spectra.} %
 \end{figure}

The polarization dependence of the Raman spectra also reveals information about degradation in the exfoliated crystals. Certain exfoliated sample spectra, particularly from samples overexposed to laser light and/or air, do not exhibit the expected selection rules. As shown in figure \ref{fig:raman}c, spectra from such samples show only a slight polarization dependence and a weak two-magnon peak. This is attributed to sample degradation and is discussed in more detail later. There are, however, important differences between bulk and exfoliated spectra even when the expected selection rules are observed. Figure \ref{fig:disc}a illustrates the evolution of the two-magnon line shape with thickness. Spectra from thin crystals ($<$ 50 nm) show a large enhancement in the amplitude of the two-magnon peak, as well as an apparent blue shift in the peak positions of the two-magnon (from around 2500 cm$^{-1}$ to 2900 cm$^{-1}$). The shift is more clearly seen in figure \ref{fig:disc}b, which shows both bulk and 30 nm spectra. It should be noted that the bulk spectrum has been scaled to match the peak heights of the exfoliated crystal. 

 \section{Discussion}
There are several factors that could potentially cause significant change in the Raman spectra of the exfoliated samples. The first is the interference correction mentioned earlier. This correction is sensitive to the thickness of the SiO$_2$ layer, so any variation from the nominal value (280 nm) would produce a distortion of the Raman spectra. The thickness of the thermal oxide layers used in this work can vary by 5 percent, according to the manufacturer and confirmed by ellipsometry measurements. Figure \ref{fig:disc}c shows the effect of this variation on the corrected spectra of a 10 unit cell exfoliated sample. The thickness of the oxide layer can affect the ratio of the two-phonon and two-magnon features as well as influence their lineshapes. However, this effect is insufficient to account for the observed blue shift and enhancement of the two-magnon. The Raman response therefore suggests a significant change in the magnetic interactions of the exfoliated crystals with respect to bulk.

A change in doping level could account for the observed change in the exfoliated crystals' effective exchange constant. One possibility is environmental doping by charged impurities on the SiO$_2$ surface. This effect has previously been observed in graphene\cite{PhysRevLett.99.246803}. Typically holes are added to the graphene layer and we might expect a similar effect in exfoliated Bi-2212. However, the two-magnon feature in hole-based cuprates is known to red-shift, weaken and broaden with doping\cite{weber_merlin,PhysRevB.68.184504}, while our spectra blue-shift and grow (see figures \ref{fig:disc}a and \ref{fig:disc}b). Another possibility is oxygen outdiffusion. This effect has been observed in Bi-2212 microwhiskers\cite{ISI:000229544200081,synchrotron} and in cuprate thin films\cite{ISI:000263564500011}. Oxygen loss would be expected to reduce the number of holes in the exfoliated crystal and so is consistent with the observed change in the two-magnon feature. Indeed, the peak position of the two-magnon feature seen in the exfoliated crystals is consistent with the peak location reported elsewhere for antiferromagnetic and insulating Bi-2212 \cite{PhysRevB.68.184504}. Loss of holes through oxygen outdiffusion is also consistent with the insulating behavior observed in optimally-doped exfoliated samples\cite{geim2005}. Lastly, one might conjecture that a structural change in the exfoliated Bi-2212 sample could be responsible for shifting the two-magnon peak. As the energy and efficiency of the two-magnon process depends on the hopping integral $t$\cite{weber_merlin,PhysRevLett.74.3057,PhysRevB.53.8619}, an increase in this parameter could explain the peak enhancement and shift observed in thin exfoliated crystals.


Finally, Bi-2212 is known to degrade in atmosphere,\cite{0953-2048-6-7-008} which would produce a corresponding change in the Raman spectra. Given the the thickness of these exfoliated samples, even thin surface degradation can be expected to have a pronounced effect on the measured Raman features. Indeed, earlier work done on exfoliated Bi-2212 crystals found that even optimally doped exfoliated crystals were insulating\cite{geim2005} when deposited in air. We find that older samples tend to show less distinct symmetry, as evidenced by the XX and X'X' spectra shown in figure \ref{fig:raman}c. Specifically, the difference between the XX and X'X' geometries is less pronounced and the overall scattered intensity is reduced. The particular sample shown in \ref{fig:raman}c was exposed to air for tens of hours as well as potentially damaged by over exposure to the Raman excitation laser. It should be noted that the loss of selection rules suggests the crystals are becoming amorphous. The two-phonon feature is also absent in \ref{fig:raman}c, consistent with a degradation of the lattice structure. Note that this degradation is distinct from the structural change discussed above as a possible cause of the enhancement and blue-shift of the two-magnon Raman feature. Raman spectroscopy therefore provides a powerful method for identifying degradation in exfoliated samples.

 \section{Conclusion}
The work described here suggests that exfoliated cuprate thin crystals can be produced with dimensions suitable for device applications. FET devices would allow the investigation of a number of important issues relating to cuprate physics, principally the disentanglement of disorder and carrier density. Significant differences were observed between the two-magnon Raman spectra of exfoliated and bulk Bi-2212, indicating a change in the effective exchange constant of the exfoliated crystals. However, when properly handled, these samples demonstrate selection rules and energy scales similar to bulk. This suggests that exfoliated crystals remain sufficiently bulk-like to merit study. Changes in the Raman spectra of these materials with time also indicate the importance of minimizing sample exposure to atmosphere and to laser radiation. The two most important directions for further research are transport measurements on exfoliated crystal devices (field effect and transport) as well as depositing Bi-2212 thin crystals on other substrates. In particular, transport measurements could help elucidate the role of disorder\cite{K.McElroy08122005} in the formation of the pseudogap and in normal state transport\cite{PhysRevLett.85.638,PhysRevB.71.014514,PhysRevLett.75.4662}. A Raman-inactive substrate would allow study of the lower energy Bi-2212 phonon modes and thus provide important insight into the structure of these cuprate thin crystals. Nonetheless, we have identified a change in the magnetic exchange constant of the exfoliated crystals with respect to bulk and have demonstrated the utility of AFM and in particular Raman microscopy in establishing the quality of these crystals.

\begin{acknowledgments}
We are grateful for numerous discussions with Y.J. Kim, A. Paramekanti, A.B. Kuzmenko, Z.Q. Li, M.Y. Han, and P. Kim. AFM was performed with the assistance of R. McAloney and M.C. Goh. We would also like to acknowledge the help of the ECTI Open Research Facility. Work at the University of Toronto was supported by NSERC, CFI, ORF and OCE. 
\end{acknowledgments}

\end{document}